\documentclass{iau} 
\usepackage{graphicx}

\title[Current driving mechanism in the BZ process]
{Where is the electric current driven in the Blandford-Znajek process?}

\author[Kenji Toma \& Fumio Takahara]
{Kenji Toma$^{1}$ \and Fumio Takahara$^2$}

\affiliation{$^1$Frontier Research Institute for Interdisciplinary Sciences, Tohoku University, Sendai 980-8578, Japan; email:{\tt toma@astr.tohoku.ac.jp} \\[\affilskip]
  $^3$Department of Earth and Space Science, Graduate School of Science, Osaka University, Toyonaka 560-0043, Japan
}

\pubyear{2016}
\volume{324}  %% insert here IAU Symposium No.
\jname{New Frontiers in Black Hole Astrophysics}
\editors{C. Mundell \& A. Gomboc, eds.}
\begin{document}

\maketitle

\begin{abstract}
The Blandford-Znajek process, the steady electromagnetic energy extraction from a rotating black hole, is widely believed to work for driving relativistic jets, although it is still under debate where the electric current is driven. We address this issue analytically by investigating the time-dependent state in the Boyer-Lindquist and Kerr-Schild coordinate systems. This analysis suggests that a non-ideal magnetohydrodynamic region is required in the time-dependent state, while not in the steady state.
\end{abstract}

\firstsection % if your document starts with a section,
\section{Introduction}

The driving mechanism of relativistic jets associated with active galactic nuclei, gamma-ray bursts, and Galactic microquasars is one of the major problems in astrophysics. A widely discussed model is based on the Blandford-Znajek (BZ) process, the energy extraction from a rotating black hole (BH) along magnetic field lines threading it (\cite[Blandford \& Znajek 1977]{bz77}). This process produces Poynting-dominated jets, in which the plasma particles can be accelerated to relativistic speeds (e.g., \cite[Komissarov et al. 2009]{komissarov09}, \cite[Lyubarsky 2009]{lyubarsky09}, \cite[Toma \& Takahara 2013]{tt13}). This theoretical picture is now being quantitatively compared with the recent observational results from the M87 jet (e.g., \cite[Asada \& Nakamura 2012]{asada12}, \cite[Kino et al. 2015]{kino15}, \cite[Porth \& Komissarov 2015]{porth15}, \cite[Mertens et al. 2016]{mertens16}).

The BZ process was proposed in the pioneering paper of \cite{bz77}, in which were found steady, axisymmetric, force-free solutions of the Kerr BH magnetosphere in the slow rotation limit where the Poynting flux is nonzero along the field lines threading the event horizon. This was followed by demonstrations of numerical and approximate analytical solutions with finite speed of BH rotation (e.g., \cite[Komissarov 2001]{komissarov01}, \cite[Koide et al. 2002]{koide02}, \cite[McKinney \& Gammie 2004]{mckinney04}, \cite[Barkov \& Komissarov 2008]{barkov08}, \cite[Tchekhovskoy et al. 2011]{tchekhovskoy11}, \cite[Ruiz et al. 2012]{ruiz12}, \cite[Contopoulos et al. 2013]{contopoulos13}, \cite[Beskin \& Zheltoukhov 2013]{beskin13}). It is also generically proved that the steady, axisymmetric situation of no Poynting flux cannot be maintained for open field lines threading the ergosphere (\cite[Toma \& Takahara 2014]{tt14}). However, the physical mechanism for the creation of the flux in the electromagnetically dominated plasma has not been clearly explained.

The issue is where the steady electric current (which produces the toroidal magnetic field) is driven in the BZ process. All the previous studies on this issue focused on the steady state condition, as far as we are aware, while we argue that this issue cannot be solved without investigating the time-dependent state towards the steady state. Then we suggest that the electric current is generally driven in the non-ideal magnetohydrodynamic (MHD) region in the time-dependent state, and such regions are hidden near (or within) the event horizon in the steady state. The steady current circuit is maintained without any current driving source, because the force-free plasma has no resistivity. In this article we explain the issue in more details and briefly summarize our argument.

\section{The issue}

\begin{figure}[t]
%% \vspace*{-2.0 cm}
\begin{center}
 \includegraphics[scale=0.35]{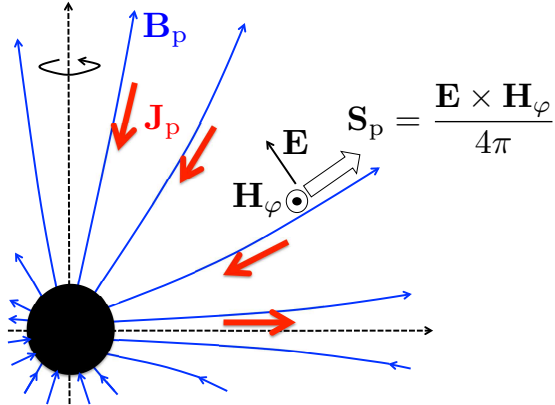} 
%% \vspace*{-1.0 cm}
 \caption{Schematic picture of the electromagnetic structure of the BZ process. }
   \label{fig1}
\end{center}
\end{figure}

The BZ process assumes that the poloidal magnetic field $\mathbf{B}_{\rm p}$ produced by the plasma filling the magnetosphere (not by the BH itself) threads the event horizon, as shown in Figure~\ref{fig1}. In the case of non-zero BH rotation, the steady axisymmetric state has the electric field $\mathbf{E}$ perpendicular to $\mathbf{B}_{\rm p}$ (with $E_\varphi = 0$) and the toroidal magnetic field $\mathbf{H}_\varphi$, which form the outward Poynting flux $\mathbf{S}_{\rm p} = \mathbf{E} \times \mathbf{H}_\varphi/4\pi$, where $\mathbf{E}$ and $\mathbf{H}$ are fields as measured in the coordinate basis. One should first note that $\mathbf{H}_\varphi$ is produced by the poloidal current $\mathbf{J}_{\rm p}$, and not produced in vacuum just by the rotation of space-time (\cite[Wald 1974]{wald74}). Second, in contrast to the BZ process, the origin of outward Poynting flux of pulsar winds has been definitely identified: The rotation of the matter-dominated central star (pulsar) generates $\mathbf{E}$ ($= -\mathbf{V}_\varphi \times \mathbf{B}_{\rm p}$) and drives $\mathbf{J}_{\rm p}$ via the unipolar induction process (\cite[Goldreich \& Julian 1969]{gj69}). We can write $\nabla \cdot \mathbf{S}_{\rm p} = -\mathbf{E} \cdot \mathbf{J}_{\rm p}$, and the source term is positive in the star, where $\mathbf{J}_{\rm p}$ is driven to flow in the direction of $-\mathbf{E}$. The BZ process, working in the electromagnetically dominated plasma, does not include any matter-dominated region in which $\mathbf{B}_{\rm p}$ is anchored. The current cannot cross the field lines, which means $\nabla \cdot \mathbf{S}_{\rm p} = 0$. Then it is unclear where $\mathbf{J}_{\rm p}$ is driven. 

The membrane paradigm (\cite[Thorne et al. 1986]{thorne86}) interprets the event horizon as a rotating conductor that generates $\mathbf{E}$ and drives $\mathbf{J}_{\rm p}$ in an analogy with the unipolar induction of pulsar winds. However, the horizon does not actively affects its exterior, but just passively absorbs particles and waves (\cite[Punsly \& Coroniti 1989]{punsly89}). The flux production has to work outside the horizon.

For such a causal flux production, some theorists proposed a scenario in which certain types of negative energies (as measured in the coordinate basis) created outside the horizon (and within the ergosphere) flow towards the horizon, resulting in the positive outward $\mathbf{S}_{\rm p}$. This is an analogous to the mechanical Penrose process. However, MHD simulations demonstrate that no regions of negative particle energy are seen in the steady state (\cite[Komissarov 2005]{komissarov05}). The concept of negative electromagnetic energy inflow (\cite[Komissarov 2009]{komissarov09b}, \cite[Lasota et al. 2014]{lasota14}, \cite[Koide \& Baba 2014]{koide14}) is not physically essential, because the sign of electromagnetic energy density ($e = -\alpha T^t_{{\rm EM},t}$, where $\alpha$ is the lapse function of Kerr metric) depends on the coordinates (\cite[Toma \& Takahara 2016]{tt16}). The total energy flux per unit mass flux treated in the analytical ideal MHD model can be negative even outside the ergosphere (\cite[Takahashi et al. 1990]{takahashi90}). The inertial drift current in the ideal MHD model can cross the field lines, but it cannot produce all of $\mathbf{S}_{\rm p}$ (\cite[Hirotani et al. 1992]{hirotani92}).

The current $\mathbf{J}_{\rm p}$ might be driven in the narrow non-ideal MHD region which steadily exists outside the horizon (\cite[Okamoto 2006]{okamoto06}, \cite[Blandford \& Znajek 1977]{bz77}). This possibility may not be excluded. If this is the case, such a non-ideal MHD region may have electric field component parallel to the magnetic field, which accelerates the particles and produces characteristic electromagnetic radiation. Thus the present issue appears to be not only theoretical but also related to observations.

\section{Our argument}

\begin{figure}[t]
%% \vspace*{-2.0 cm}
\begin{center}
 \includegraphics[scale=0.42]{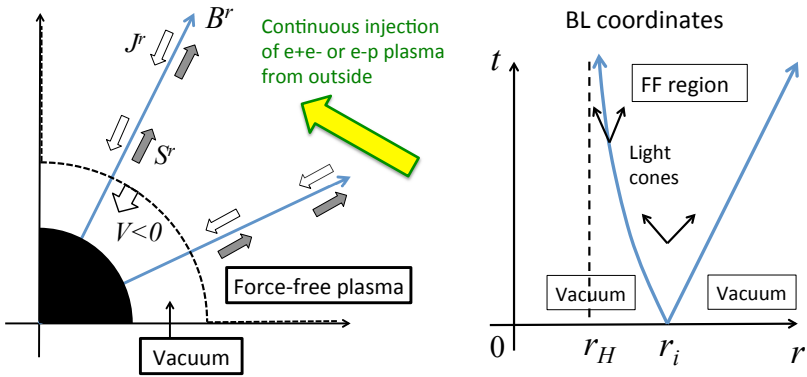} 
%% \vspace*{-1.0 cm}
 \caption{Our toy model of the time-dependent state (focusing on the inflow region) (\textit{left}) and its schematic space-time diagram in the Boyer-Lindquist coordinates (\textit{right}).}
   \label{fig2}
\end{center}
\end{figure}

To help understanding the present issue, let us consider how the Schwarzschild gravitational field is formed. Its origin is definitely the mass inside the horizon. However, the information that the mass exists inside the horizon never comes out to the exterior. Then how is the Schwarzschild gravitational field formed? This question cannot be properly answered only by considering the static state. The answer is that this gravitational field has been produced by the mass before it falls across the horizon.

This consideration suggests that the origin of steady electromagnetic field surrounding the horizon cannot either be understood without investigating the time-dependent state towards the steady state. All the previous studies on the current driving mechanism (as far as we are aware) investigated only the steady state.

Although such a time-dependent state should be analyzed numerically, we use a toy model to qualitatively illustrate the process of building the poloidal current circuit (\cite[Toma \& Takahara 2016]{tt16}). In order to find the essential physics, our analysis is performed in the Boyer-Lindquist (singular at the horizon) and Kerr-Schild (regular at the horizon) coordinate systems. As shown in Figure~\ref{fig2} (left), we initially consider a vacuum in Kerr space-time, and then begin the continuous injection of force-free plasma particles between the two light surfaces as a Gedankenexperiment. The inflow (outflow) fills the vacuum near the horizon (at infinity). Then we can see a process building the poloidal current circuit. Our toy model assumes that $\mathbf{B}_{\rm p}$ is fixed to be split-monopole in the whole region, that the two boundaries between the force-free region and the vacuum are geometrically thin and move radially, and that the force-free region and the vacuum have their steady-state structures, but the values of the physical quantities ($E$, $H_\varphi$, etc.) keep being updated as determined by the varying conditions of the inner and outer boundaries. These assumptions enable us to integrate the $3+1$ Maxwell equations. Then we found that $\mathbf{J}_{\rm p}$ has to flow on the boundaries (in the directions for making the circuit). More importantly, we found that the displacement current at the inner boundary also regulates $\mathbf{J}_{\rm p}$ in the force-free region, i.e., $H_\varphi^{\rm ff} = \sqrt{\gamma} V (D_{\rm ff}^\theta - D_{\rm vac}^\theta) - 4\pi \sqrt{\gamma} \eta^\theta$, where the subscript and superscript ``ff" and ``vac" denote the force-free and vacuum regions, respectively, and $\mathbf{D}$, $V$, $\eta^\theta$, and $\gamma$ are the electric field as measured by the zero angular momentum observers, the velocity of inner boundary, the surface current on the inner boundary, and the determinant of spatial metric, respectively.

The outward Poynting flux in the force-free region is produced at the inner boundary as $\nabla \cdot \mathbf{S}_{\rm p} = -\partial_t e - \mathbf{E} \cdot \mathbf{J}_{\rm p}$. This process takes place outside the horizon, which does not violate causality. Figure~\ref{fig2} (right) is the schematic space-time diagram of the two boundaries in the Boyer-Lindquist coordinates. When the inner boundary approaches the horizon, the outward signal from it propagates slower and slower and it can hardly affect the force-free region. This will lead to the steady state, and one has $\nabla \cdot \mathbf{S}_{\rm p} = 0$, i.e., there is no driving source of $\mathbf{J}_{\rm p}$ in the steady state. Since the force-free plasma is assumed to have no resistivity, the current can keep flowing steadily. 

In conclusion, the driving source of $\mathbf{J}_{\rm p}$ or the non-ideal MHD region exists outside the horizon in the time-dependent state, while it is not required in the steady state. See \cite{tt16} for more details.

\end{document}